\documentclass{PoS}

\usepackage{graphicx}
\usepackage{grffile}
\usepackage{amsmath}

\title{The low-lying spectrum of $\mathcal{N}=1$ supersymmetric Yang-Mills theory}

\ShortTitle{The low-lying spectrum of $\mathcal{N}=1$ supersymmetric Yang-Mills theory}

\author{\speaker{Pietro Giudice}, Gernot M\"unster, Stefano Piemonte\\
  Universit\"at M\"unster, Institut f\"ur Theoretische Physik, \\
  Wilhelm-Klemm-Str. 9, D-48149 M\"unster, Germany\\
  E-mail: \email{p.giudice@uni-muenster.de, munsteg@uni-muenster.de, 
spiemonte@uni-muenster.de}}

\author{Georg Bergner\\
  Universit\"at Bern, Institut f\"ur Theoretische Physik, \\
  Sidlerstr.~5, CH-3012 Bern, Switzerland\\
  E-mail: \email{bergner@itp.unibe.ch}
}

\author{Istvan Montvay\\
  Deutsches Elektronen-Synchrotron DESY, \\
  Notkestr. 85, D-22603 Hamburg, Germany\\
  E-mail: \email{montvay@mail.desy.de}
}

\abstract{
The spectrum of the lightest bound states in $\mathcal{N} = 1$ supersymmetric 
Yang-Mills theory with SU(2) gauge group, calculated on the lattice, is 
presented. The masses have first been extrapolated towards vanishing gluino 
mass and then to the continuum limit. The final picture is consistent with 
the formation of degenerate supermultiplets.
}

\FullConference{The European Physical Society Conference on High Energy Physics\\
		 22-29 July 2015\\
		 Vienna, Austria}

\newcommand{\aetap}{\text{a--}\eta'}
\newcommand{\api}{{\text{a--}\pi}}
\newcommand{\afn}{\text{a--}f_0}

\newcommand{\tr}[1]{\ensuremath{\mathrm{Tr}} \left[{#1}\right]}
\newcommand{\I}{\ensuremath{\mathrm{i}\hspace{1pt}}}

\newcommand{\beq}{\begin{equation}}
\newcommand{\eeq}{\end{equation}}
\newcommand{\bea}{\begin{eqnarray}}
\newcommand{\eea}{\end{eqnarray}}

\begin{document}

\section{Introduction}
Supersymmetry (SUSY) was first studied in the early seventies and since
then it has been a central topic in particle physics. 
Fermions, the constituents of matter, have previously been considered 
fundamentally different from bosons,
the particles that transmit the forces between them. In SUSY, fermions and  
bosons are unified.

The only way we have to study SUSY non-perturbatively from first principles
is by lattice field theory approach: the theory is studied
on a space-time that has been discretised onto a lattice.

The simplest non-abelian supersymmetric gauge theory, which is the subject 
of our study, is the $\mathcal{N} = 1$ supersymmetric 
Yang-Mills (SYM) theory with gauge group SU(2).  It describes the 
interaction between gluons and gluinos.  
The theory is asymptotically free at high energies and becomes strongly
coupled in the infrared limit.  The mass spectrum 
is expected to consist of  colourless  bound  states.   
If  supersymmetry  is  unbroken, particles  should  belong  to  mass
degenerate SUSY multiplets.

The investigation of the mass spectrum of the theory is intrinsically 
a non-perturbative problem.
First predictions~\cite{Veneziano:1982ah,Farrar:1997fn} were only possible 
constraining the form of the low-energy effective actions by means of the 
symmetries of the theory.
Verifying these predictions is the central task of our work.
Some important results have already been obtained by our collaboration in 
previous studies, see~ \cite{Demmouche:2010sf,Bergner:2012rv,Bergner:2013nwa}.
The theory has also been studied at non-zero temperatures 
in~\cite{Bergner:2014saa}.

\section{$\mathcal{N} = 1$ SYM theory}
The Euclidean on-shell action for $\mathcal{N} = 1$ SYM theory in the
continuum is given by:
\beq
S(g,m_g) =  \int d^4 x \left\{\frac{1}{4} (F_{\mu\nu}^a F_{\mu\nu}^a) + \frac{1}{2}\bar{\lambda}_a (\gamma^\mu D^{ab}_\mu + m)\lambda_b - \frac{\Theta}{16\pi}\epsilon_{\mu\nu\rho\sigma}F^{\mu\nu}F^{\rho\sigma}\right\} \ .
\label{eq:action1}
\eeq
The theory contains gluons ($A_\mu(x)$ gauge fields) as bosonic particles, 
and gluinos ($\lambda(x)$ fields) as their fermionic superpartners.
It is an extension of pure gauge theory: SU($N_c$) gauge symmetry is imposed
together with a single conserved supercharge, obeying the algebra:
\beq
\{ Q_\alpha,Q_\beta \}=(\gamma^\mu C)_{\alpha, \beta} P_\mu \ .
\label{commutationrelation}
\eeq
The Majorana spinors $Q_\alpha$ are the generators of the supersymmetry,
$C$ is the charge conjugation matrix and $P_\mu$ the momentum operator.
In our present investigations the gauge group is SU(2), {\it i.~e.}~$N_c=2$.
Supersymmetry relates gluons and gluinos:
\bea
A_\mu(x) & \rightarrow & A_\mu(x) -2 \,\I \bar{\lambda}(x)\gamma_\mu \epsilon \\
\lambda^a(x) & \rightarrow & \lambda^a(x) - \sigma_{\mu\nu} F^a_{\mu\nu}(x) \epsilon \ ,
\eea
where $\epsilon$ is a global fermionic parameter (consistent with the 
Majorana condition), parametrising the transformation.

In Eq.~\ref{eq:action1} the $\Theta$-term has been added to the action 
as can also be done for QCD.
It does not violate the underlying symmetries of the model. 
In out study only the case $\Theta = 0$ has been considered. The role
of this term to study the topological properties of the theory can be seen
in~\cite{Bergner:2014ska}.

\section{$\mathcal{N} = 1$ SYM theory on the lattice}
\label{sec:symonthelattice}
In our simulations, the gauge part of the action is discretised with a 
tree-level Symanzik improved action, while the gluino part is represented 
using a discretised version of the Dirac operator, which depends on the 
gauge fields in the adjoint representation, which is known as Wilson-Dirac 
operator.
The latter depends on the hopping parameter $\kappa$
which is inversely proportional to the gluino mass.
To reduce the lattice artifacts also in the fermionic part of the action, 
we apply one level of stout smearing to the gauge fields in the Wilson-Dirac 
operator~\cite{Bergner:2012rv}.
The parameter $m$, which introduces a bare mass for the gluino, breaks 
supersymmetry softly, {\it i.~e.} the main features of the supersymmetric 
theory, concerning the ultraviolet renormalisability, remain intact.

There are two other sources of supersymmetric breaking:
the finite volume effects~\cite{Bergner:2012rv} and the discretisation 
of the theory on the space-time lattice~\cite{Bergner:2009vg}.
The former have been already clarified and it is totally under control in 
our simulations. 
Above a box size of about $1.2$~fm (in QCD units) the statistical errors 
and the systematic errors of the finite size effects, with our current 
numerical precision, are of the same order 
and hence the finite size effects can be neglected.
The latter is instead a fundamental problem related to the
breaking of translational invariance on the lattice. Consequently, 
the relation~\ref{commutationrelation} cannot be satisfied on the lattice.

In~\cite{Bergner:2009vg} a No-Go theorem is discussed:
a realisation of a complete supersymmetry on the lattice
can only be achieved with nonlocal interactions and a nonlocal derivative.
This statement is circumvented by fine tuning of the bare parameters 
of the theory towards the supersymmetric continuum limit.
In SYM the fine-tuning of a single parameter is enough
to approach not only supersymmetry but also the
(spontaneously broken) chiral symmetry~\cite{Curci:1986sm,Suzuki:2012pc}:
the hopping parameter $\kappa$ is tuned to the point where the renormalised 
gluino mass vanishes. In practice, this is achieved by monitoring the mass 
of the unphysical adjoint pion $\api$, which is defined by 
the connected contribution of the $\aetap$ correlator. 
A formal definition can be given in a partially quenched 
setup~\cite{Munster:2014cja}. An important result, which was also 
established by means of the OZI approximation~\cite{Veneziano:1982ah},
is that  the adjoint pion mass vanishes for a massless gluino
with the law: $m^2_\api \propto m_g$.

\section{Setting the scale}
Monte Carlo simulations are able to extract only dimensionless quantities.
The lattice spacing, which characterises the system under simulation,
is not an input parameter.
To make contact with the measurements done in a laboratory, the value
of the lattice spacing $a$ has to be determined. This value 
is implicitly defined once a dimensionful observable, for instance 
a distance in physical units $d$, is chosen as a reference to set the scale.
Matching the result of a simulation $\hat{d}=d/a$, with the previous value, one
determines the desired quantity: $a=d/\hat{d}$.

In the last years mainly two different observables have been chosen
to set the scale, both of which have the dimension of a length: 
the Sommer parameter $r_0$~\cite{Sommer:1993ce} and 
the Wilson flow scale $w_0$~\cite{Borsanyi:2012zs}.

$r_0$ is defined as the distance $r$ such that the strong force between a 
static quark-antiquark pair, multiplied by the squared distance, $r^2 F(r)$,
is equal to a specific value, typically 1.65. Systematic errors arise
because of different smoothing procedures that can be used to improve 
the signal of $F(r)$ and because of different fitting procedures employed to
extract the parameters.

The scale $w_0$ is based on the gradient flow generated by the Wilson or 
the Symanzik gauge field action. The procedure is specified by a partial 
differential equation, similar to a diffusion equation, which evolves the 
gauge field along a fictitious time. $w_0$, as we show also in this work, 
is less effected by systematic errors and provides a more reliable method 
to set the scale.

\section{Mass spectrum from effective Lagrangians}
\label{sec:sprectrummodels}
An important similarity between Yang-Mills theory and 
 \emph{supersymmetric}  Yang-Mills theory is confinement.
In the supersymmetric theory the glueballs are completed 
with the mesonic states,
the gluinoballs, and the fermionic gluino-glueballs. 
The low-lying spectrum of particles has been predicted by 
means of effective Lagrangians.
\begin{figure}[t]
\hspace{-5mm}
\includegraphics[width=7.9cm]{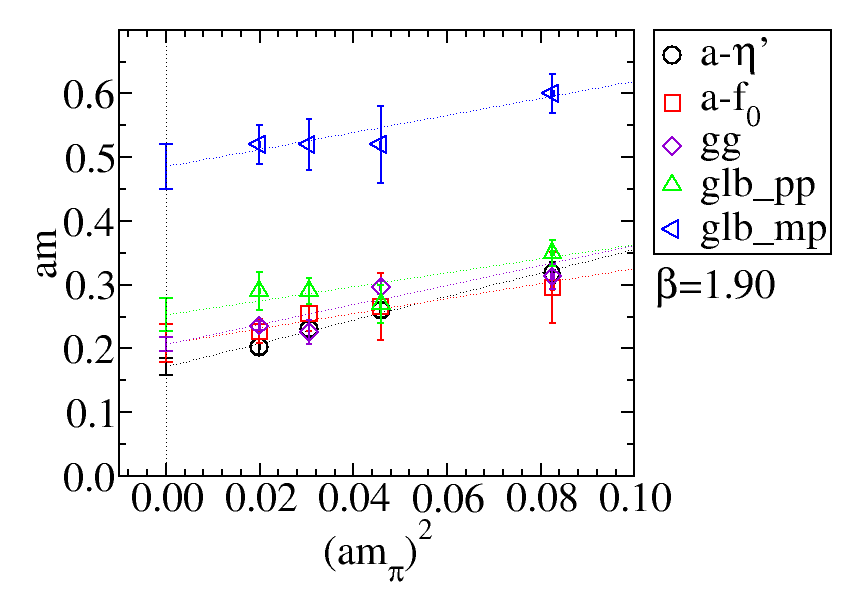}
\includegraphics[width=7.9cm]{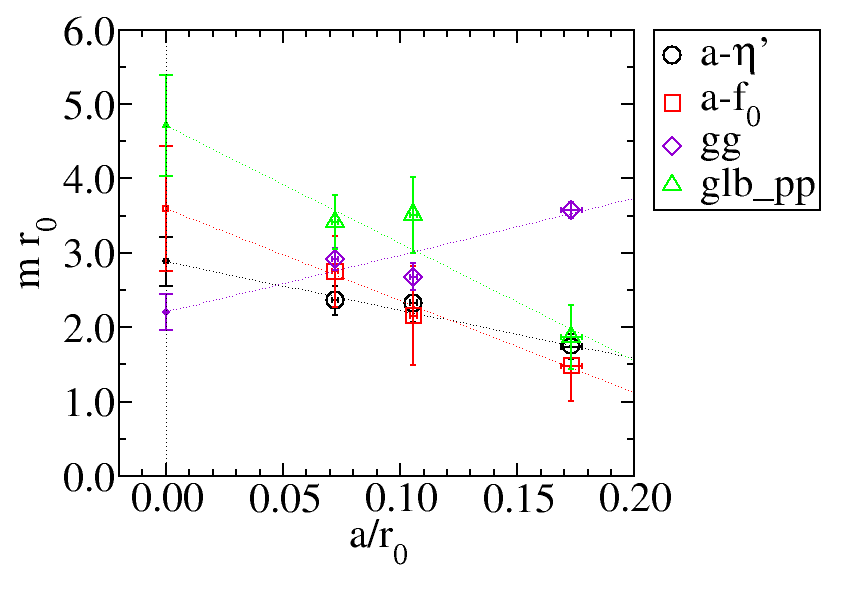}
\caption{\textbf{(Left)} Mass spectrum (gg: gluino-glue; glb$\_$pp: glueball $0^{++}$; 
  glb$\_$mp: glueball $0^{-+}$) at $\beta=1.90$ for four values of 
  the $\api$ mass. The extrapolated chiral limit is shown as well.
  \textbf{(Right)} Mass spectrum extrapolated to the continuum limit, using $r_0$.}
\label{fig:spectrum}
\end{figure}
In~\cite{Veneziano:1982ah} a first supermultiplet was described:
it consists of a scalar ($0^+$ gluinoball: $\afn \sim \bar\lambda \lambda$),
a pseudoscalar ($0^-$ gluinoball: $\aetap \sim \bar\lambda \gamma_5 \lambda$),
and a Majorana fermion (spin $1/2$ gluino-glueball:
$\chi \sim \sigma^{\mu \nu} \tr{ F_{\mu \nu} \lambda }  $).
This first work was generalised in~\cite{Farrar:1997fn} introducing 
pure gluonic states in the effective Lagrangian.
In addition to the first chiral supermultiplet a new one was found:
a $0^-$ glueball, a $0^+$ glueball, and again a gluino-glueball.

It is worth noting that neither of these supermultiplets contains
a pure gluino-gluino, a gluino-gluon or a gluon-gluon bound state:
the physical excitations are mixed states of them.

In~\cite{Farrar:1997fn} it has been argued that the glueball states, in 
the supersymmetric limit, are lighter than the gluinoball states.
Later, other authors~\cite{Feo:2004mr}, using different arguments,
and information about ordinary QCD, deduce that the lighter states
are gluinoballs. This point has not been totally clarified so far.

\section{Mass spectrum from Monte Carlo simulations}
The most important signature for the existence of a continuum limit 
with unbroken supersymmetry is the low-lying mass spectrum of bound states:
we expect in fact degeneracy inside the same supermultiplet.

The investigation of the spectrum of the theory represents a non-perturbative 
challenge, which requires numerical simulations of the theory, discretised on 
a space-time lattice, with the help of supercomputers.
One of the tasks of our project is to show that the results of the numerical 
simulations are consistent with restoration of supersymmetry in the 
continuum limit. We have already published results on the spectrum 
of the theory for two values of the lattice spacing, which is characterised 
by the value of the parameter $\beta$:
in~\cite{Demmouche:2010sf} we presented our results at $\beta=1.60$; 
in \cite{Bergner:2013nwa,Bergner:2013jia} we decreased the value of
the lattice spacing by $\sim 40\%$ using $\beta=1.75$. In the latter,
for the first time we had indication of restoration of SUSY.
In~\cite{Bergner:2014iea} we presented our preliminary results at $\beta=1.90$,
where the lattice spacing was further reduced by $\sim 30\%$.
In this conference proceedings we present our final results at $\beta=1.90$
for the first time, with the extrapolation to the continuum 
limit of the spectrum. In Figure~\ref{fig:spectrum} (Left) we plot five
bounds states for different values of the adjoint pion mass squared;
they are linearly extrapolated to the chiral limit, according to 
Section~\ref{sec:symonthelattice}. 
This numerical extrapolation shows a degeneracy of the masses in the lowest 
supermultiplet within two standard deviations. Note that the glueball $0^{++}$
is particularly noisy: it is well known that this channel requires a statistic
$\sim 20$ times larger than the others to have comparable errors. In our
case all channels are measured on the same sample of field configurations.
\FIGURE{
  \includegraphics[width=7.9cm]{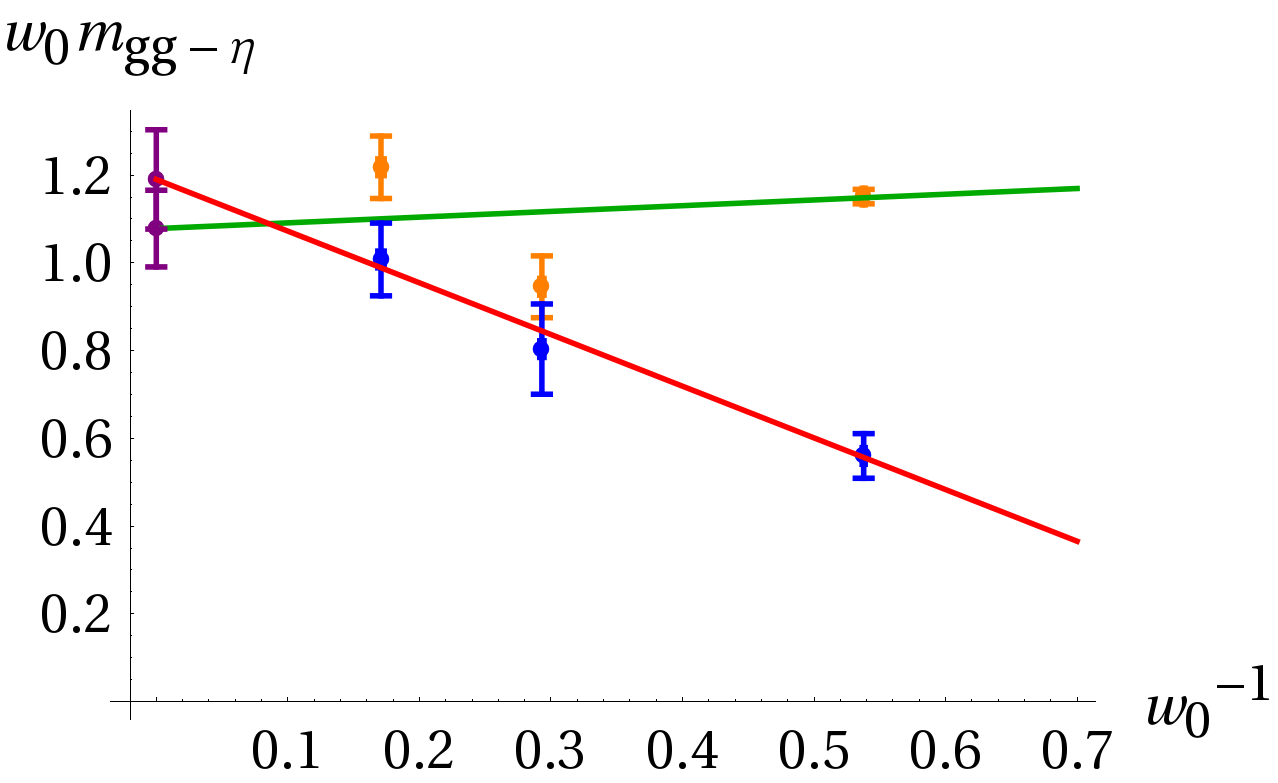}
  \caption{Extrapolation to the continuum limit of the gluino-glueball $\chi$ 
and the gluinoball $\aetap$, using $w_0$.}
  \label{fig:continuum-w0}
}
The lattice methods used in this work allow to extract directly 
only of the lowest mass in each channel. 
The operator used to extract the mass of the glueball  $0^{++}$ has a strong 
mixing with $\afn$, indeed they have the same quantum numbers, and therefore
the value of its mass fall in the lower supermultiplet. 
The operator used to describe the glueball $0^{-+}$ instead has a weak
mixing with the $\aetap$.
As a consequence in Figure~\ref{fig:spectrum} (Left) we see
that only the glueball $0^{-+}$ is present in the higher supermultiplet.
These facts seem to suggest that the glueball states are higher than the 
gluinoball as argued in~\cite{Feo:2004mr}.

The spectrum of the theory extrapolated to the continuum limit using
the inverse of the Sommer scale $r_0$ is shown in 
Figure~\ref{fig:spectrum} (Right). According to this plot we finally 
can state that we see the restoration, within two standard deviation, 
of supersymmetry in the continuum limit. 

In Figure~\ref{fig:continuum-w0} we present for comparison the extrapolation
to the continuum limit of two particles which belong to the same
supermultiplet, the gluino-glueball $\chi$ and the gluinoball $\aetap$.
In this case, using the scale parameter $w_0$, we clearly see that in the 
continuum limit the two particles are degenerate. In general, we have
verified that the extrapolations based on the parameter  $w_0$ give
better results.

\section{Conclusion and Outlooks}
We have studied the spectrum of $\mathcal{N} = 1$ supersymmetric Yang-Mills 
theory non-perturbatively employing lattice calculations.
We have shown that, within two standard deviations, the theory is
characterised by degeneracy of the mass spectrum inside the same supermultiplet.
In other words, it has been possible to see restoration of supersymmetry 
in the continuum limit.

Our next step is to clarify the role of the mixing, studying
the excited states of the theory. Moreover we are planning to move from 
SU(2) to the more realistic SU(3) gauge theory, which is part of the 
supersymmetric extension of the Standard Model.

\section*{Acknowledgements}
This project is supported by the German Science Foundation (DFG) under
contract Mu 757/16.
The authors gratefully acknowledge the computing time granted by the John
von Neumann Institute for Computing (NIC) and provided on the supercomputers
JUQUEEN and
JUROPA at J\"ulich Supercomputing Centre (JSC).
Further computing time has been
provided by the compute cluster PALMA of the University of M\"unster.


\end{document}